\documentclass[fleqn,twoside]{article}
\usepackage{espcrc2}

\usepackage{graphicx}
\usepackage[figuresright]{rotating}

\def\10{$SO(10)$}
\def\21{SU(2) $\otimes$ U(1) }

\def\422{$SU(4) \otimes SU(2) \otimes SU(2)$}
\def\321{SU(3) $\otimes$ SU(2) $\otimes$ U(1)}
\def \nbb {$\beta\beta_{0\nu}$ }

\def\lsim{\raise0.3ex\hbox{$\;<$\kern-0.75em\raise-1.1ex\hbox{$\sim\;$}}}
\def\gsim{\raise0.3ex\hbox{$\;>$\kern-0.75em\raise-1.1ex\hbox{$\sim\;$}}}

\DeclareMathAlphabet{\mathsc}{OT1}{cmr}{m}{sc}

\def\21{$SU(2) \otimes U(1) $}

\newcommand{\flux}[2][]{\ensuremath{\ifthenelse{\equal{#1}{}}{}{^{#1}\!}\mathit{#2}}}

\newcommand{\CL}   {C.L.}
\newcommand{\dof}  {d.o.f.}
\newcommand{\eVq}  {\mathrm{eV}^2}
\newcommand{\Sol}  {\textsc{sol}}

\newcommand{\Atm}  {\textsc{atm}}

\newcommand{\Dms}  {\Delta m^2_\Sol}
\newcommand{\Dma}  {\Delta m^2_\Atm}

\newcommand{\Lsnd} {\textsc{lsnd}}
\newcommand{\Dml}  {\Delta m^2_\Lsnd}

\newcommand{\AddrAHEP}{%
  AHEP Group, Instituto de F\'{\i}sica Corpuscular --
  C.S.I.C./Universitat de Val{\`e}ncia \\
  Edificio Institutos de Paterna, Apt 22085, E--46071 Valencia, Spain}

\title{Physics of Massive Neutrinos}

\author{J.~W.~F.~Valle\address[MCSD]\AddrAHEP
  \thanks{Work supported by grant BFM2002-00345, by EC RTN grant
    HPRN-CT-2000-00148 and ESF \emph{Neutrino Astrophysics Network}.
    Latest neutrino oscillation plots are taken from a review with
    Maltoni, Schwetz and Tortola.}  }
       
\begin{document}

\begin{abstract}
  I summarize the present status of global analyses of neutrino
  oscillations, including the most recent KamLAND and K2K data, as
  well as the latest solar and atmospheric neutrino fluxes.  I give
  the allowed ranges of the three--flavour oscillation parameters from
  the current worlds' global neutrino data sample, their best fit
  values and discuss the small parameters $\alpha \equiv \Dms/\Dma$
  and $\sin^2\theta_{13}$, which characterize the strength of CP
  violation in neutrino oscillations.
  I briefly discuss neutrinoless double beta decay and the LSND
  neutrino oscillation hint, as well as the robustness of the neutrino
  oscillation results in the presence of non-standard physics.

  \vspace{1pc}
\end{abstract}

\maketitle

\section{INTRODUCTION}

The discovery of neutrino oscillations by combining data from
solar~\cite{solarNF}, atmospheric~\cite{atmNF},
reactor~\cite{reacNu04} and accelerator~\cite{accelNu04} neutrino
studies has marked a turning point in our understanding of nature and
has brought neutrino physics to the center of attention of the
particle, nuclear and astrophysics communities. This culminates a
heroic effort which started over four decades ago and now firmly
establishes the existence of small neutrino masses, confirming
theoretical expectations which date back to the early eighties.
Neutrino mass arise from the dimension-five operator $\ell \ell \phi
\phi$ where $\phi$ the \21 Higgs doublet and $\ell$ is a lepton
doublet~\cite{Weinberg:1980bf}.  Nothing is known from first
principles about the mechanism that induces this operator, its
associated mass scale or flavour structure.  Its most popular
realization is the seesaw mechanism which induces small neutrino
masses from the exchange of heavy states. Although inspired by
unification, the effective model-independent description of the seesaw
at low-energies is in terms of the \21 gauge structure and contains a
small effective Higgs triplet contribution to the neutrino
mass~\cite{schechter:1980gr}. Such general seesaw has 24 parameters, 3
more than the tripletless case~\cite{schechter:1980gr}~\footnote{In
  the mass basis these correspond to the 12 mixing angles and 12 CP
  phases (both Dirac and Majorana-type) that characterize the full
  3$\times$6 charged current seesaw lepton mixing
  matrix~\cite{schechter:1980gr}.  These are exactly the same
  parameters involved in the description of
  leptogenesis~\cite{Fukugita:1986hr}, though in this case the use of
  the weak basis seems more convenient. Note also that current
  nomenclature (type-I versus II seesaw) is opposite from the original
  one in \cite{schechter:1980gr}.}.

The structure of the three-flavour lepton mixing matrix in various
gauge theories of neutrino mass was given in~\cite{schechter:1980gr}
and it was also argued that, on \emph{general} grounds, massive
neutrinos should be Majorana particles, leading to neutrinoless double
beta decay (\nbb)~\cite{Wolfenstein:1981rk}.  Current neutrino
oscillation data are well described by the simplest unitary lepton
mixing matrix neglecting CP violation. The effect of Dirac CP phases
in oscillations and Majorana phases in \nbb constitute the main
challenge for the years to come.


Here I focus mainly on the determination of neutrino mass and mixing
parameters in neutrino oscillation studies, a currently thriving
industry~\cite{industry}, with many new experiments currently underway
or planned. The interpretation of the data requires good calculations
of solar and atmospheric neutrino
fluxes~\cite{Bahcall:2004fg,Honda:2004yz}, neutrino cross sections and
experimental response functions, as well as a careful description of
neutrino propagation both in the Sun and the Earth, including matter
effects~\cite{mikheev:1985gs,wolfenstein:1978ue}.  After summarizing
the latest (post-Neutrino-2004) global analysis of 3-neutrino
oscillation parameters~\cite{Maltoni:2004ei}~\cite{Goswami2004} I
briefly discuss their impact on future \nbb
searches~\cite{Bilenky:2004wn}.
This writeup combines my plenary talk and the WG1 parallel session
talk on non-standard scenarios of neutrino conversion.

\section{TWO-NEUTRINO OSCILLATIONS}

Let us first consider neutrino oscillations in the two-flavour
approximation~\cite{Maltoni:2004ei}.

\subsection{Solar + KamLAND}
\label{sec:solar-+-kamland}

The solar neutrino data includes the measured rates of the chlorine
experiment at the Homestake mine ($2.56 \pm 0.16 \pm 0.16$~SNU), the
most up-to-date results of the gallium experiments SAGE
($66.9~^{+3.9}_{-3.8}~^{+3.6}_{-3.2}$~SNU) and GALLEX/GNO ($69.3 \pm
4.1 \pm 3.6$~SNU), as well as the 1496--day Super-K data in the form
of 44 bins (8 energy bins, 6 of which are further divided into 7
zenith angle bins). The SNO data include the most recent data from the
salt phase in the form of the neutral current (NC), charged current
(CC) and elastic scattering (ES) fluxes, as well as the 2002 spectral
day/night data (17 energy bins for each day and night period).

The analysis methods are described in~\cite{Maltoni:2003da} and
references therein~\cite{Goswami2004}. We use a generalization of the
pull approach for the $\chi^2$ calculation originally suggested in
Ref.~\cite{Fogli:2002pt} in which all systematic uncertainties such as
those of the eight solar neutrino fluxes are included by introducing
new parameters in the fit and adding a penalty function to the
$\chi^2$.  The method \cite{Maltoni:2004ei,Maltoni:2003da} is exact to
all orders in the pulls and covers the case of correlated statistical
errors~\cite{Balantekin:2003jm} as necessary to treat the SNO--salt
experiment.  This is particularly interesting as it allows us to
include the Standard Solar Model $^8$B flux prediction as well as the
SNO NC measurement on the same footing, without pre-selecting a
particular value, as implied by expanding around the predicted value:
the fit itself chooses the best compromise between the SNO NC data and
the SSM prediction.

KamLAND detects reactor anti-neutrinos at the Kamiokande site by the
process $\bar\nu_e + p \to e^+ + n$, where the delayed coincidence of
the prompt energy from the positron and a characteristic gamma from
the neutron capture allows an efficient reduction of backgrounds.
Most of the incident $\bar{\nu}_e$ flux comes from nuclear plants at
distances of $80-350$ km from the detector, far enough to probe the
LMA solution of the solar neutrino problem.
The neutrino energy is related to the prompt energy by $E_\nu =
E_\mathrm{pr} + \Delta - m_e$, where $\Delta$ is the neutron-proton
mass difference and $m_e$ is the positron mass.
For lower energies there is a relevant contribution from geo-neutrino
events to the signal~\cite{Fiorentini:2003ww}. To avoid large
uncertainties associated with the geo-neutrino flux an energy cut at
2.6~MeV prompt energy is applied for the oscillation analysis.

First KamLAND results were taken from March to October 2002.  Data
corresponding to a 162 ton-year exposure gave 54 anti-neutrino events
in the final sample, after all cuts, while $86.8 \pm 5.6$ events are
predicted for no oscillations with $0.95\pm 0.99$ background events.
The probability that the KamLAND result is consistent with the
no--disappearance hypothesis is less than 0.05\%.  This gave the first
evidence for the disappearance of neutrinos traveling to a detector
from a power reactor and the first terrestrial confirmation of the
solar neutrino anomaly.

New KamLAND data were presented at Neutrino 2004~\cite{reacNu04}.
With a somewhat larger fiducial volume of the detector an exposure
corresponding to 766.3~ton-year has been obtained between March 2002
and January 2004 (including a reanalysis of the previous 2002 data).
In total 258 events have been observed, versus $356.2\pm 23.7$ reactor
neutrino events expected in the case of no disappearance and $7.5\pm
1.3$ background events. This leads to a confidence level of 99.995\%
for $\bar\nu_e$ disappearance, and the averaged survival probability
is $0.686 \pm 0.044\mathrm{(stat)} \pm 0.045\mathrm{(syst)}$. Moreover
evidence for spectral distortion consistent with oscillations is
obtained~\cite{reacNu04}.

It is convenient to treat the latest KamLAND data binning them equally
in $1/E_\mathrm{pr}$, instead of the traditional bins of equal size in
$E_\mathrm{pr}$. Various systematic errors associated to the neutrino
fluxes, backgrounds, reactor fuel composition and individual reactor
powers, small matter effects, and improved $\bar{\nu}_e$ flux
parameterization are included~\cite{Maltoni:2004ei}.  One finds a
beautiful agreement between KamLAND data and the region implied by the
LMA solution to the solar neutrino problem, which in this way has been
singled out as the only viable one in contrast to the previous ``zoo''
of oscillation solutions~\cite{Maltoni:2003da,gonzalez-garcia:2000sq}.
However the stronger evidence for spectral distortion in the recent
data leads to improved information on $\Dms$, substantially reducing
the allowed region of oscillation parameters. From this point of view
KamLAND has played a fundamental role in the resolution of the solar
neutrino problem.

\begin{figure}[htb]
\includegraphics[height=5cm,width=0.9\linewidth]{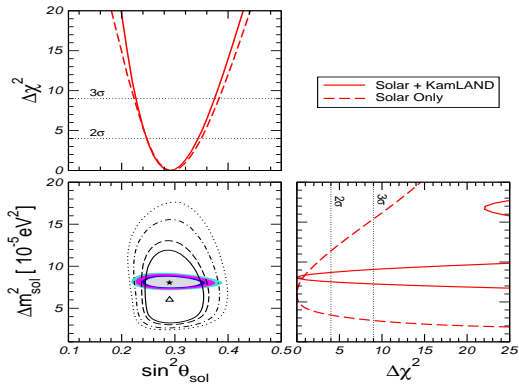}
\caption{\label{fig:solkam-region} %
  Regions allowed by solar+KamLAND data at 90\%, 95\%, 99\%, and
  3$\sigma$ \CL\ for 2 \dof. Unshaded regions correspond to solar data
  only.}
\end{figure}

Assuming CPT invariance one can directly compare the information
obtained from solar neutrino experiments with the KamLAND reactor
results.
In Fig.~\ref{fig:solkam-region} we show the allowed regions from the
combined analysis of solar and KamLAND data.  

\subsection{Atmospheric + K2K}
\label{sec:atmospheric-+-k2k}

The zenith angle dependence of the $\mu$-like atmospheric neutrino
data from the Super-K experiment provided the first evidence for
neutrino oscillations in 1998, an effect confirmed also by other
atmospheric neutrino experiments~\cite{atmNF}.  The dip in the $L/E$
distribution of the atmospheric $\nu_\mu$ survival probability seen in
Super-K gives a clearer signature for neutrino oscillations.

\begin{figure}[t] \centering
\vspace{9pt}
    \includegraphics[height=5cm,width=0.9\linewidth]{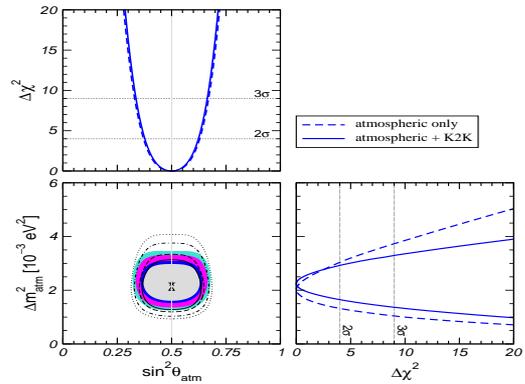}
    \caption{\label{fig:atm+k2k} %
      $\sin^2\theta_\Atm$--$\Dma$ regions allowed at 90\%, 95\%, 99\%,
      and 3$\sigma$ \CL\ for 2 \dof\ (the unshaded regions include
      atmospheric data only). }
\end{figure}

The analysis summarized below includes the most recent charged-current
atmospheric neutrino data from Super-K, with the $e$-like and
$\mu$-like data samples of sub- and multi-GeV contained events grouped
into 10 zenith-angle bins, with 5 angular bins of stopping muons and
10 through-going bins of up-going muon events.  No information on
$\nu_\tau$ appearance, multi-ring $\mu$ and neutral-current events is
used since an efficient Monte-Carlo simulation of these data would
require more details of the Super-K experiment, in particular of the
way the neutral-current signal is extracted from the data (see
Refs.~\cite{Maltoni:2003da,gonzalez-garcia:2000sq} for details). 
Here the new three--dimensional atmospheric neutrino fluxes given in
~\cite{Honda:2004yz} are used, in contrast to previous analyses using
the Bartol fluxes~\cite{barr:1989ru}.  With this one obtains the
regions of two-flavour $\nu_\mu\to\nu_\tau$ oscillation parameters
$\sin^2\theta_\Atm$ and $\Dma$ shown by the hollow contours in
Fig.~\ref{fig:atm+k2k}. One notes that the $\Dma$ values obtained with
the three--dimensional atmospheric neutrino fluxes are lower than
obtained previously~\cite{Maltoni:2003da}, in excellent agreement with
the results of the Super-K collaboration~\cite{hayato:2003}.

The KEK to Kamioka (K2K) long-baseline neutrino oscillation
experiment~\cite{accelNu04} probes $\nu_\mu$ disappearance in the same
$\Delta m^2$ region as probed with atmospheric neutrinos. The neutrino
beam is produced by a 12~GeV proton beam from the KEK proton
synchrotron, and consists of 98\% muon neutrinos with a mean energy of
1.3~GeV. The beam is controlled by a near detector 300~m away from the
proton target. Information on neutrino oscillations is obtained by the
comparing this near detector data with the $\nu_\mu$ content of the
beam observed by the Super-Kamiokande detector at a distance of
250~km.

The data K2K-I sample has been collected in the period from June 1999
to July 2001 ($4.8\times 10^{19}$ protons on target) and gave 56
events in Super-K, whereas $80.1^{+6.2}_{-5.4}$ were expected for no
oscillations. The probability that the observed flux is explained by a
statistical fluctuation without neutrino oscillations is less than
1\%~\cite{accelNu04}.  K2K-II started in Fall 2002, and released data
at the Neutrino2004 conference~\cite{accelNu04} corresponding to
$4.1\times 10^{19}$ protons on target, comparable to the K2K-I sample.
Altogether K2K-I and K2K-II give 108 events in Super-K, to be compared
with $150.9^{+11.6}_{-10.0}$ expected for no oscillations. Out of the
108 events 56 are so-called single-ring muon events.  This data sample
contains mainly muon events from the quasi-elastic scattering $\nu_\mu
+ p \to \mu + n$, and the reconstructed energy is closely related to
the true neutrino energy.  The K2K collaboration finds that the
observed spectrum is consistent with the one expected for no
oscillation only at a probability of 0.11\%, whereas the spectrum
predicted by the best fit oscillation parameters has a probability of
52\%~\cite{accelNu04}.

The re-analysis of K2K data given in \cite{Maltoni:2004ei} uses the
energy spectrum of the 56 single-ring muon events from K2K-I + K2K-II
(unfortunately not the full K2K data sample of 108 events, for lack of
information outside the K2K collaboration). Under reasonable
assumptions, one can fit the data divided into 15 bins in
reconstructed neutrino energy.
One finds that the neutrino mass-squared difference indicated by the
$\nu_\mu$ disappearance in K2K agrees with atmospheric neutrino
results, providing the first confirmation of oscillations with $\Dma$
from a man-made neutrino source. However K2K gives a rather weak
constraint on the mixing angle due to low statistics in the current
data sample.

The shaded regions in Fig.~\ref{fig:atm+k2k} are the allowed
($\sin^2\theta_\Atm$,~$\Dma$) regions that follow from the combined
analysis of K2K and Super-K atmospheric neutrino data.
One sees that, although the determination of $\sin^2\theta_\Atm$ is
completely dominated by atmospheric data, the K2K data start already
to constrain the allowed region of $\Dma$.
Note also that, despite the downward shift of $\Dma$ implied by the
new atmospheric fluxes, the new result is statistically compatible
both with the previous one in~\cite{Maltoni:2003da} and with the value
obtained by the Super-K $L/E$ analysis~\cite{atmNF}.  Note that the
K2K constraint on $\Dma$ from below is important for future
long-baseline experiments, as such experiments are drastically
affected if $\Dma$ lies in the lower part of the 3$\sigma$ range
indicated by the atmospheric data alone.

\section{THREE-NEUTRINO OSCILLATIONS}

The effective leptonic mixing matrix in gauge theories of massive
neutrinos was systematically studied in~\cite{schechter:1980gr}.  For
high-scale seesaw models this matrix is approximately unitary and, for
three neutrinos, may be taken as~\cite{hagiwara:2002fs}~\footnote{This
  also holds in radiative models~\cite{zee:1980ai,babu:1988ki} and
  models where supersymmetry is the origin of neutrino
  mass~\cite{Hirsch:2004he}. In inverse seesaw
  models~\cite{mohapatra:1986bd} deviations from unitarity may be
  phenomenologically important~\cite{bernabeu:1987gr}.}
\begin{equation}
  \label{eq:2227}
K =  \omega_{23} \omega_{13} \omega_{12}  
\end{equation}
where each factor, for example,
$$\omega_{13} = \left(\begin{array}{ccccc}
c_{13} & 0 & e^{i \phi_{13}} s_{13} \\
0 & 1 & 0 \\
-e^{-i \phi_{13}} s_{13} & 0 & c_{13} 
\end{array}\right)\,.
$$
contains an angle and a CP phase. Two of the three angles are
involved in solar and atmospheric oscillations, so we set $\theta_{12}
\equiv \theta_\Sol$ and $\theta_{23} \equiv \theta_\Atm$. All these
three phases are physical~\cite{schechter:1981gk}, one corresponds to
the one present in the quark sector (Dirac-phase) and affects neutrino
oscillations, while the other two are associated to the Majorana
nature of neutrinos and show up in neutrinoless double beta decay and
other lepton-number violating processes, but not in conventional
neutrino oscillations~\cite{schechter:1981gk,doi:1981yb}.

Current neutrino oscillation experiments are insensitive to CP
violation, thus all phases will be neglected~(future neutrino
factories aim at probing the effects of the Dirac
phase~\cite{Dick:1999ed}). In this approximation three-neutrino
oscillations depend on the three mixing parameters $\sin^2\theta_{12},
\sin^2\theta_{23}, \sin^2\theta_{13}$ and on the two mass-squared
differences $\Dms \equiv \Delta m^2_{21} \equiv m^2_2 - m^2_1$ and
$\Dma \equiv \Delta m^2_{31} \equiv m^2_3 - m^2_1$ characterizing
solar and atmospheric neutrinos.  The hierarchy $\Dms \ll \Dma$
implies that one can set, to a good approximation, $\Dms = 0$ in the
analysis of atmospheric and K2K data, and $\Dma$ to infinity in the
analysis of solar and KamLAND data.
The relevant neutrino oscillation data in a global three-neutrino
analysis are those of sections \ref{sec:solar-+-kamland} and
\ref{sec:atmospheric-+-k2k} together with the constraints from the
CHOOZ and Palo Verde reactor experiments~\cite{apollonio:1999ae}.

\begin{figure}[t] \centering
    \includegraphics[width=.95\linewidth]{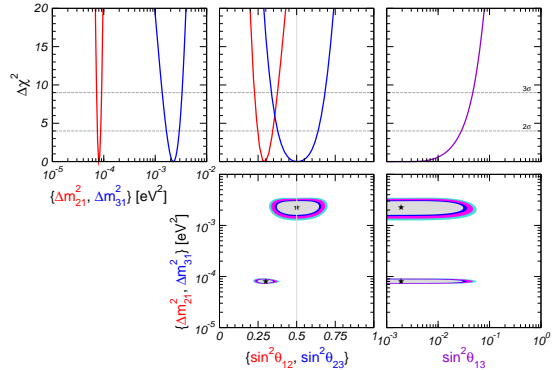}
    \caption{\label{fig:global} %
      Three--neutrino regions allowed by the world's oscillation data
      at 90\%, 95\%, 99\%, and 3$\sigma$ \CL\ for 2 \dof\. Top panels
      give $\Delta \chi^2$ minimized with respect to all undisplayed
      parameters.}
\end{figure}

The results of the global three--neutrino analysis are summarized in
Fig.~\ref{fig:global} and in Tab.~\ref{tab:summary}. In the upper
panels of the figure the $\Delta \chi^2$ is shown as a function of the
parameters $\sin^2\theta_{12}, \sin^2\theta_{23}, \sin^2\theta_{13},
\Delta m^2_{21}, \Delta m^2_{31}$, minimized with respect to the
undisplayed parameters. The lower panels show two-dimensional
projections of the allowed regions in the five-dimensional parameter
space. The best fit values and the allowed 3$\sigma$ ranges of the
oscillation parameters from the global data are summarized in
Tab.~\ref{tab:summary}.  This table gives the current status of the
three--flavour neutrino oscillation parameters.
\begin{table}[t] \centering    \catcode`?=\active \def?{\hphantom{0}}
      \begin{tabular}{|l|c|c|}        \hline        parameter & best
      fit & 3$\sigma$ range         \\        \hline        $\Delta
      m^2_{21}\: [10^{-5}~\eVq]$        & 8.1?? & 7.2--9.1 \\
      $\Delta m^2_{31}\: [10^{-3}~\eVq]$        & 2.2?? &  1.4--3.3 \\
      $\sin^2\theta_{12}$        & 0.30? & 0.23--0.38 \\
      $\sin^2\theta_{23}$        & 0.50? & 0.34--0.68 \\
      $\sin^2\theta_{13}$        & 0.000 & $\leq$ 0.047 \\
      \hline    
\end{tabular}    \vspace{2mm} 
\caption{\label{tab:summary} Three--neutrino oscillation parameters 
from the global data analysis given in \cite{Maltoni:2004ei}.}
\end{table}
  
In a three--neutrino scheme CP violation disappears when two neutrinos
become degenerate~\cite{schechter:1980gr} or when one angle vanihes,
$\theta_{13} \to 0$~\cite{schechter:1980bn}. Genuine three--flavour
effects are associated to the mass hierarchy parameter $\alpha \equiv
\Dms/\Dma$ and the mixing angle $\theta_{13}$.
\begin{figure}[t] \centering
    \includegraphics[height=3.5cm,width=.95\linewidth]{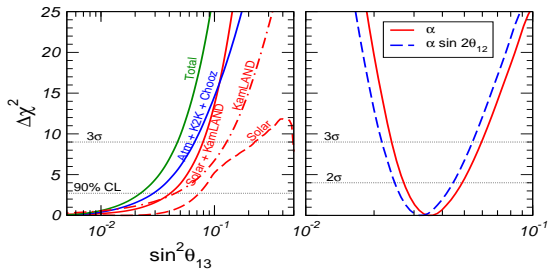}
    \caption{\label{fig:alpha}%
      Determination of $\alpha \equiv \Dms / \Dma$ and bound on
      $\sin^2\theta_{13}$ from current neutrino oscillation data.}
\end{figure}
The left panel in Fig.~\ref{fig:alpha} gives the parameter $\alpha$ as
determined from the global $\chi^2$ analysis of \cite{Maltoni:2004ei}.
The figure also gives $\Delta\chi^2$ as a function of the parameter
combination $\alpha \sin 2\theta_{12}$ which, to leading order,
determines the long baseline $\nu_e\to\nu_\mu$ oscillation
probability~\cite{Freund:2001pn,Akhmedov:2004ny}. The last unknown
angle in the three--neutrino leptonic mixing matrix is $\theta_{13}$,
for which only an upper bound exists.
The leftt panel in Fig.~\ref{fig:alpha} gives $\Delta\chi^2$ as a
function of $\sin^2\theta_{13}$ for different data sample choices.
One finds that the new data from KamLAND have a surprisingly strong
impact on this bound. Before the KamLAND-2004 data the bound on
$\sin^2\theta_{13}$ from global data was dominated by the CHOOZ
reactor experiment, together with the determination of $\Delta
m^2_{31}$ from atmospheric data.  However, with the KamLAND-2004 data
the bound becomes comparable to the reactor bound, and contribute
significantly to the final global bound 0.022~(0.047) at 90\% \CL\ 
(3$\sigma$) for 1 \dof\ This improved $\sin^2\theta_{13}$ bound
follows from the strong spectral distortion found in the 2004
sample~\cite{Maltoni:2004ei}.
\begin{figure}[t] \centering
    \includegraphics[height=4cm,width=.75\linewidth]{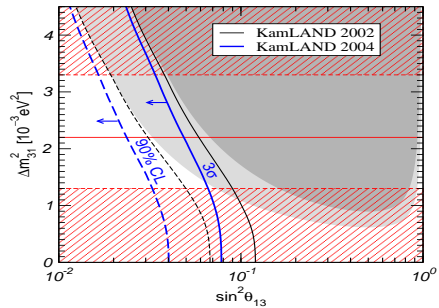}
    \caption{\label{fig:t13-solar-chooz} Upper bound on
      $\sin^2\theta_{13}$ (1 \dof) from solar+KamLAND+CHOOZ data
      versus $\Dma$. Dashed (solid) curves correspond to 90\%
      (3$\sigma$) \CL\ bounds, thick curves include KamLAND-2004 data,
      thin ones do not.  Light (dark) regions are excluded by CHOOZ at
      90\% (3$\sigma$) \CL\ The horizontal line corresponds to the
      current $\Dma$ best fit value, hatched regions are excluded
      by atmospheric + K2K data at 3$\sigma$.}
\end{figure}
Note that, since the CHOOZ bound on $\sin^2\theta_{13}$ deteriorates
quickly as $\Dma$ decreases (see Fig.~\ref{fig:t13-solar-chooz}), the
improvement is especially important for lower $\Dma$ values, as
implied by the new three--dimensional atmospheric
fluxes~\cite{Honda:2004yz}.
In Fig.~\ref{fig:t13-solar-chooz} we show the upper bound on
$\sin^2\theta_{13}$ as a function of $\Dma$ from CHOOZ data alone
compared to the bound from an analysis including solar and reactor
neutrino data. One finds that, although for larger $\Dma$ values the
bound on $\sin^2\theta_{13}$ is dominated by CHOOZ, for $\Dma \lsim 2
\times 10^{-3} \eVq$ the solar + KamLAND data start being relevant.
The bounds implied by the 2002 and 2004 KamLAND data are compared in
Fig.~\ref{fig:t13-solar-chooz}. In addition to reactor and accelerator
neutrino oscillation searches~(Lindners' talk), future studies of the
day/night effect in large water Cerenkov solar neutrino experiments
like UNO or Hyper-K~\cite{SKatm04} may improve the sensitivity on
$\sin^2\theta_{13}$~\cite{Akhmedov:2004rq}.

\section{ABSOLUTE NEUTRINO MASSES}

\begin{figure}[b]
  \centering
\includegraphics[width=5cm,height=3cm]{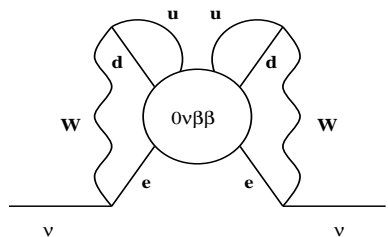}  
  \caption{Equivalence between \nbb and Majorana mass in gauge theories~\cite{Schechter:1981bd}.}
 \label{fig:bbox} 
\end{figure}

Neutrino oscillation data are sensitive only to mass differences, not
to the absolute neutrino masses.  Nor do they have any bearing on the
fundamental issue of whether neutrinos are Dirac or Majorana
particles~\cite{schechter:1981gk,doi:1981yb}.  The significance of the
\nbb decay is given by the fact that, in a gauge theory, irrespective
of the mechanism that induces \nbb, it is bound to also yield a
Majorana neutrino mass~\cite{Schechter:1981bd}, as illustrated in Fig.
\ref{fig:bbox}.
Quantitative implications of the ``black-box'' argument are
model-dependent, but the theorem holds in any ``natural'' gauge
theory.

Now that oscillations are experimentally confirmed we know that \nbb
must be induced by the exchange of light Majorana neutrinos. The
corresponding amplitude is sensitive both to the absolute scale of
neutrino mass as well as the two Majorana CP phases that characterize
the minimal 3-neutrino mixing matrix~\cite{schechter:1980gr}.
\begin{figure}[t]
  \centering
\includegraphics[width=.7\linewidth,height=5cm]{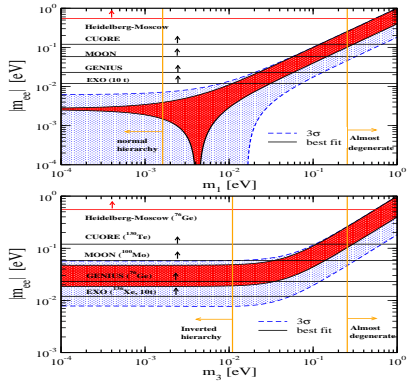}
  \caption{Neutrino-mass-induced \nbb from current oscillation data 
    versus current and projected experimental sensitivities.}
 \label{fig:nbbfut} 
\end{figure}
Fig. \ref{fig:nbbfut} shows the estimated average mass parameter
characterizing the neutrino exchange contribution to \nbb versus the
lightest neutrino mass.  The upper (lower) panel corresponds to the
cases of normal (inverted) neutrino mass spectra.  The calculation
takes into account the current neutrino oscillation parameters from
\cite{Maltoni:2004ei} and the nuclear matriz elements
of~\cite{Bilenky:2004wn}.
In contrast to the normal hierarchy, where a destructive interference
of neutrino amplitudes is possible, the inverted neutrino mass
hierarchy implies a ``lower'' bound for the \nbb amplitude.
Quasi-degenerate neutrinos~\cite{caldwell:1993kn,Ioannisian:1994nx}
such as predicted in \cite{babu:2002dz}, give the largest \nbb
amplitude. Future experiments~\cite{Klapdor-Kleingrothaus:1999hk} will
provide an independent confirmation of the Heidelberg-Moscow
results~\cite{Klapdor-Kleingrothaus:2004wj} and extend the sensitivity
to inverse hierarchy models.

Complementary information on the absolute scale of neutrino mass comes
from beta decays searches~\cite{Osipowicz:2001sq} as well as
cosmology~\cite{Hannestad:2004nb}.

\section{FOUR-NEUTRINO OSCILLATIONS?}

In addition to the strong evidence for oscillations of solar and
atmospheric neutrinos there is also a hint for oscillations from
LSND~\cite{aguilar:2001ty}. This accelerator experiment at Los Alamos
observed $87.9\pm22.4 \pm 6.0$ excess events in the
$\bar\nu_\mu\to\bar\nu_e$ appearance channel, corresponding to a
transition probability of $P=(0.264\pm0.067\pm0.045)\%$, which is
$\sim 3.3~\sigma$ away from zero. To explain this signal with neutrino
oscillations requires a mass-squared difference $\Dml \sim 1~\eVq$, a
value inconsistent with the solar+KamLAND and atmospheric+K2K
experiments described above.  Four--neutrino
schemes~\cite{caldwell:1993kn,peltoniemi:1993ec,peltoniemi:1993ss}
where a sterile neutrino is added to the three active ones might
provide the additional mass scale needed to reconcile the LSND
evidence.

The updated analysis described in \cite{Maltoni:2004ei} includes, in
addition to LSND, solar+KamLAND and atmospheric+K2K data, also the
data from short-baseline (SBL) accelerator and reactor experiments
with no evidence for oscillations.  The oscillation data are divided
into the four sets $X$ = SOL, ATM, NEV, LSND and the PG method is used
to test their statistical compatibility in a given mass scheme.
Following Ref.~\cite{maltoni:2002xd} we consider the contributions to
$\Delta\chi^2_X = \chi^2_X - (\chi^2_X)_\mathrm{min}$ from different
data samples (see~\cite{maltoni:2001bc} for more details). In
Tab.~\ref{tab:pg} we show the contributions of the 4 data sets to
$\chi^2_\mathrm{PG} \equiv \bar\chi^2_\mathrm{min}$ for (3+1) and
(2+2) oscillation schemes. As expected we note that in (3+1) schemes
the main contribution comes from SBL data due to the tension between
LSND and NEV data in these schemes.  For (2+2) schemes a large part of
$\chi^2_\mathrm{PG}$ comes from solar and atmospheric data, due to the
rejection against a sterile neutrino contribution of these two data
sets. The contribution from NEV data in (2+2) comes mainly from the
tension between LSND and KARMEN~\cite{Church:2002tc}, which does not
depend on the mass scheme.
\begin{table*}[t]\centering
    \catcode`?=\active \def?{\hphantom{0}}
    \begin{tabular}{|c|cccc|c|c|}
        \hline
        & SOL & ATM & LSND & NEV &   $\chi^2_\mathrm{PG}$ & PG \\
        \hline
        (3+1) & 0.0 & ?0.4 & 5.7 & 10.9 & 17.0 & $1.9 \times 10^{-3} \: (3.1\sigma)$ \\
        (2+2) & 5.3 & 20.8 & 0.6 & ?7.3 & 33.9 & $7.8 \times 10^{-7} \: (4.9\sigma)$ \\
        \hline
    \end{tabular}
    \caption{Contributions of different data sets to
      $\chi^2_\mathrm{PG}$ in (3+1) and (2+2) neutrino mass schemes.}
    \label{tab:pg}
\end{table*}
The parameter goodness of fit is obtained by evaluating
$\chi^2_\mathrm{PG}$ for 4 \dof~\cite{Maltoni:2003cu}. 
Using an improved goodness of fit method especially sensitive to the
combination of data sets one finds that (2+2) schemes are ruled out at
the $4.9\sigma$ level. The inclusion of MACRO data in the analysis
would further improve the degree of rejection of these schemes.
Note that such a strong rejection of (2+2) schemes holds irrespective
of whether LSND is confirmed or not.

Although none of the four--neutrino schemes provides a good fit to the
global oscillation data including LSND, it is interesting to consider
the \textit{relative} status of the three hypotheses (3+1), (2+2) and
the three--active neutrino scenario (3+0).  This is done by comparing
the $\chi^2$ value of the best fit point -- which occurs for the (3+1)
scheme -- with the ones corresponding to (2+2) and (3+0).  First we
observe that (2+2) schemes are strongly disfavored with respect to
(3+1) with a $\Delta \chi^2 = 16.9$. For 4 \dof\ this is equivalent to
an exclusion at $3.1\sigma$.
Furthermore, one finds that (3+0) is disfavored with a $\Delta \chi^2 =
17.5$ (corresponding to $3.2\sigma$ for 4 \dof) with respect to (3+1),
reflecting the high statistical significance of the LSND result.

On the other hand (3+1) spectra are disfavored by the disagreement of
LSND with short-baseline disappearance data of CDHS and Bugey, leading
to a marginal GOF of $1.9\times 10^{-3}$ ($3.1\sigma$). Should LSND be
confirmed a positive signal is predicted right at the edge of the
current sensitivity.
\begin{figure}[t] \centering
    \includegraphics[height=4cm,width=0.9\linewidth]{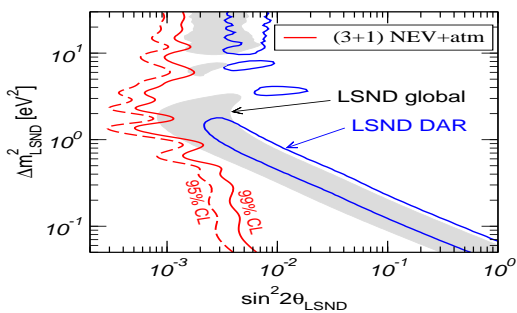}
    \caption{\label{fig:3+1} %
      Upper bound on $\sin^22\theta_\Lsnd$ from NEV, atmospheric and
      K2K data in (3+1) schemes. The dotted line corresponds to the
      99\%~\CL\ bound without K2K and using the 1-d Bartol flux.  Also
      shown are the regions allowed at 99\% \CL\ (2~\dof) from global
      LSND data and decay-at-rest (DAR) LSND data.}
\end{figure}
In Fig.~\ref{fig:3+1} we give the upper bound on the LSND oscillation
amplitude $\sin^22\theta_\Lsnd$ from the combined analysis of NEV and
atmospheric neutrino data.
From the figure one sees that the bound is incompatible with the
signal observed in LSND at the 95\%~\CL\ Marginal overlap regions
between the bound and global LSND data exist only if both are taken at
99\%~\CL\ Restricting to the decay-at-rest LSND data
sample~\cite{Church:2002tc} makes the disagreement even more severe.
This shows that 4-neutrino descriptions of LSND do not provide a
satisfactory fit to the world's neutrino data. Although a reasonable
5-neutrino fit is possible, the confirmation of LSND is essential.

\section{BEYOND  OSCILLATIONS}

Non-standard physics may in principle affect neutrino propagation
properties and detection cross sections~\cite{pakvasa:2003zv}.
Such interactions (NSI, for short) are a natural outcome of many
neutrino mass models~\cite{valle:1991pk} and can be of two types:
flavour-changing (FC) and non-universal (NU).
They may arise from a nontrivial structure of charged and neutral
current weak interactions with non-unitary lepton mixing
matrix~\cite{schechter:1980gr} as expected in inverse seesaw
models~\cite{mohapatra:1986bd,bernabeu:1987gr}, in radiative neutrino
mass models~\cite{zee:1980ai,babu:1988ki} and in supersymmetric
unified models~\cite{hall:1986dx}.
\begin{figure}[t] \centering
    \includegraphics[height=2cm,width=.6\linewidth]{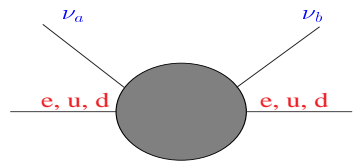}
    \caption{\label{fig:nuNSI} %
      Flavour-changing effective operator.}
\end{figure}
NSI may be schematically represented as effective dimension-6 terms of
the sub-weak strength $\varepsilon G_F$, see Fig.~\ref{fig:nuNSI}.

In the presence of NSI, the Hamiltonian describing atmospheric neutrino
propagation has, in addition to the standard oscillation part, another
term $H_\mathrm{NSI}$ accounting for an effective potential induced by
the NSI with matter:
\begin{equation}
    H_\mathrm{NSI} = \pm \sqrt{2} G_F N_f
    \left( \begin{array}{cc}
        0 & \varepsilon \\ \varepsilon & \varepsilon'
    \end{array}\right) \,.
\end{equation}
Here $+(-)$ holds for neutrinos (anti-neutrinos) and $\varepsilon$ and
$\varepsilon'$ parameterize the NSI: $\sqrt{2} G_F N_f \varepsilon$ is
the forward scattering amplitude for the FC process $\nu_\mu + f \to
\nu_\tau + f$ and $\sqrt{2} G_F N_f \varepsilon'$ represents the
difference between $\nu_\mu + f$ and $\nu_\tau + f$ elastic forward
scattering. Here $N_f$ is the number density of the fermion $f$ along
the neutrino path.

The impact of non-standard neutrino interactions on the determination
of atmospheric neutrino parameters $\Dma$ and $\sin^2\theta_\Atm$ was
considered in Ref.~\cite{fornengo:2001pm} treating the NSI strengths
as free phenomenological parameters and taking for $f$ the down-type
quark, for definiteness.  This analysis takes into account both the
effect of $\nu_\mu \to \nu_\tau$ oscillations as well as the existence
of non-standard neutrino--matter interactions in this channel.  It is
shown that, in the 2-neutrino approximation, the determination of the
atmospheric neutrino oscillation parameters $\Dma$ and $\sin^2
2\theta_\Atm$ is practically unaffected by the presence of NSI. Future
neutrino factories will substantially improve this
bound~\cite{huber:2001zw}. In contrast, as shown in
\cite{Miranda:2004nb} the determination of solar neutrino parameters
is still not yet fully robust against the presence of NSI.
\begin{figure}[t] \centering
    \includegraphics[width=2.8cm,height=\linewidth,angle=-90]{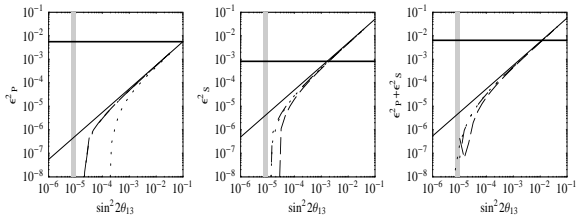}
    \caption{\label{fig:nuf13} %
      Deterioration of $\theta_{13}$ at a neutrino factory in the
      presence of NSI, explanation in \cite{huber:2002bi}}
\end{figure}
Before concluding, let me mention that, even a small residual
non-standard interaction of neutrinos in the $e-\tau$ channel leads to
a drastic loss in sensitivity in the determination $\theta_{13}$
through the so-called ``golden channels'' at a neutrino
factory~\cite{huber:2002bi}, see Fig.~\ref{fig:nuf13}. This may only
be partly overcome by combining if different baselines. Only by
rejecting NSI at a near detector can one improve sensitivities in
$\theta_{13}$.

\section{CONCLUSION}

Experiment is ahead of theory in neutrino physics: despite the great
progress achieved recently we are very far from a ``road map'' to the
ultimate theory of neutrino properties.  We have no idea of the
underlying neutrino mass generation mechanism, its characteristic
scale or its flavor structure. We have still a long way to go!


\end{document}